\documentclass[runningheads,a4paper]{llncs}
\usepackage{amssymb}
\usepackage{graphicx}
\usepackage{url}
\usepackage{graphicx,amssymb,float,makeidx,amsmath,pstricks,epsf}
\usepackage{cite}
\urldef{\mailsa}\path|aitmessaoudlysa@gmail.com|
\urldef{\mailsb}\path|fmerazka@usthb.dz|

\newcommand{\keywords}[1]{\par\addvspace\baselineskip
\noindent\keywordname\enspace\ignorespaces#1}

\begin{document}

\mainmatter  
\title{PSO and CPSO Based Interference Alignment for $K$-User MIMO Interference Channel}
\titlerunning{PSO and CPSO Based IA for $K$-User MIMO IC}
\author{Lysa AIT MESSAOUD \and Fatiha MERAZKA}%

\institute{LISIC Lab. Telecommunication department, Electronic and Computer Science Faculty,\\
USTHB University, BP. 32, EL-Alia Bab Ezzouar 16111, Algiers, Algeria.\\
\mailsa, \mailsb\\}
\maketitle

\begin{abstract}
This paper investigates how to use a metaheuristic based technique, namely Particle Swarm Optimization (PSO), in carrying out of Interference Alignment (IA) for $K$-User MIMO Interference Channel (IC). Despite its increasing popularity, mainly in wireless communications, IA lacks of explicit and straightforward design procedures. Indeed, IA design results in complex optimization tasks involving a large amount of decision variables, together with a problem of convergence of the IA solutions. In this paper the IA optimization is performed using PSO and Cooperative PSO (CPSO) more suitable for large scale optimization, a comparison between the two versions is also carried out. This approach seems to be promising.
\end{abstract}
\section{INTRODUCTION}

Multiple Input Multiple Output (MIMO) systems \cite{P1} allow improving throughput and/or reliability in wireless communications. More Precisely, a $K$-user MIMO Interference Channel (MIMO IC) models a network of $K$ transmit-receiver pairs where each transmitter communicates multiple data streams to its respective receiver generating interference at all other receivers,  which limits the whole spectral efficiency. Mitigating this interference is of major concern, and much research has been carried out to this purpose.

Interference Alignment (IA) is promising interference management technique, the idea is to divide, at each receiver, the space spanned by the available signaling dimensions into two subspaces of suitable dimensions, interference will be aligned in one subspace, and the second one will serve to contain the desired signal \cite{P2}. IA solutions are iterative since closed form solutions are only available for certain low-dimensional configurations of MIMO IC.
IA iterative schemes result in complex optimization tasks involving a large amount of decision variables, together with a problem of convergence. If this convergence is not reached, only suboptimal solution can be found at the expense of high computational complexity increases rapidly with the size of the MIMO IC (number of users, antennas or data streams).

 Many iterative IA solutions have been developed using usual numerical optimization methods, like alternating optimization \cite{P4}, Steepest Descent (SD) \cite{P5} and Gauss-Newton (GN) \cite{P3} algorithms. For more details, a complete a comparative study of optimization algorithms dedicated to MIMO IC is available in \cite{P6}.

On the other hand, Particle Swarm Optimization (PSO) is a stochastic population-based optimization technique, introduced by Kennedy and Eberhart in 1995 \cite{PSO95}. Known for combining simplicity and efficiency, PSO has been successfully applied to a wide range of engineering problems. However, it is well known that most stochastic optimization algorithms suffer from the so-called "curse of dimensionality," which simply means that their performance deteriorates as the dimensionality of the search space increases \cite{CPSO}. More precisely, it has been found in \cite{CSO.bis} that PSO perform poorly when the optimization problem is high dimensional, this results a premature convergence. To alleviate this drawback, Potter suggested in \cite{Potter} that the search space should be partitioned by splitting the solution vectors into smaller vectors, then each of these smaller search spaces is searched by a separate mechanism. This cooperative approach is said Cooperative Coevolution (CC). In \cite{CPSO}, authors applied Potter's technique to the PSO, resulting in new cooperative PSO models. In this paper, both the PSO algorithm and its cooperative version are applied to achieve IA. 

This paper is organized as follows: Section II states the IA problem. Section III gives an overview of the PSO and CPO, then describes the proposed solution. Simulation results are presented in Section IV. The paper is concluded in Section V.
\section{PROBLEM STATEMENT}
Consider a $K$-user MIMO IC consisting of $K$ transmitter receiver pairs equipped with $M_i$ and $N_i$ antennas, $i=1, \cdots,K$, respectively. Each transmitter sends $d_i$ data streams to its corresponding receiver. Using the same notation as in \cite{P3}, the system is expressed as $\prod_{i=1}^{K}\left(M_i \times N_i,d_i \right)$. The signal at each receiver is given by
\begin{equation}
\begin{aligned}
\mathbf{z}_i=& \mathbf{U}_i^H\mathbf{H}_{ii}\mathbf{V}_i\mathbf{s}_i+\sum_{i\neq j}\mathbf{U}_i^H\mathbf{H}_{ij}\mathbf{V}_j\mathbf{s}_j+\mathbf{n}_i,\\
& i=1, \cdots,K
\end{aligned}
\end{equation}
where $\mathbf{U}_i\in \mathbb{C}^{N_i \times d_i}$ and $\mathbf{V}_i\in \mathbb{C}^{M_i \times d_i}$ are the decoding and precoding matrices, respectively; $\mathbf{H}_{ij} \in \mathbb{C}^{N_i \times M_j}$ is the channel coefficients matrix between transmitter $i$ and receiver $j$, $\mathbf{s}_i$ are the symbols transmitted by user $i$ and $\mathbf{n}_i$ is the additive white Gaussian noise at the $i$th receiver. In order to perform IA, the decoding and precoding matrices must be calculated so as to fulfill the following equations \cite{P3}
\begin{equation}
\mathbf{U}_i^H\mathbf{H}_{ij}\mathbf{V}_j=0,~\forall i\neq j
\label{eq:P3.2}	
\end{equation}
\begin{equation}
\text{rank} \left(\mathbf{U}_i^H\mathbf{H}_{ii}\mathbf{V}_i\right)=d_i,~\forall i.
\label{eq:P3.3}	
\end{equation}

Condition (\ref{eq:P3.3}) is almost surely satisfied if the channel matrices $\mathbf{H}_{ij}$ do not have any special structure and both $\mathbf{U}_i$ and $\mathbf{V}_j$ are full column rank \cite{P3, P3bis}. This is verified in the calculations performed in this study, since the channel matrices $\mathbf{H}_{ij}$ are generated randomly.

Let $\mathbf{x}$ the vector containing all the optimization variables, that is, the variables in $\mathbf{U}_i$ and $\mathbf{V}_j$ as $\mathbf{x}=[\text{vec}\left(\mathbf{V}_1\right)^T,\cdots,\text{vec}\left(\mathbf{V}_K\right)^T,\text{vec}\left(\mathbf{U}_1^H\right)^T, \cdots,$ $ \text{vec}\left(\mathbf{U}_K^H\right)^T]^T$, where $\text{vec}(\mathbf{A})$ denotes the vector obtained by stacking the columns of matrix $\mathbf{A}$ below one another. Consequently, $\mathbf{x}$ contains the totality of $N_v=\sum_i(M_i+N_i)d_i$  variables in the system. Define as $\mathbf{r}(\mathbf{x})$ the function evaluating the residuals of the equations in (\ref{eq:P3.2}) which consists of $N_e=\sum_{i\neq j}d_id_j$ scalar equations, i.e., $\mathbf{r}(\mathbf{x})=[\mathbf{r}_ {21}^T, \cdots, \mathbf{r}_ {(K-1)K}^T]^T$, where $\mathbf{r}_{ij}=\text{vec}(\mathbf{U}_i^H\mathbf{H}_{ij}\mathbf{V}_j)$, to be feasible, the system $\mathbf{r}:\mathbb{C}^{N_v} \rightarrow \mathbb{C}^{N_e}$ must verify $N_v \geq  N_e$. Finally, in order to obtain a mono objective optimization problem, authors in \cite{P3} express the cost function, also called Interference Leakage (IL), as
\begin{equation}
f(\mathbf{x})=\mathbf{r}(\mathbf{x})^H\mathbf{r}(\mathbf{x}):	\mathbb{C}^{N_v}\rightarrow \mathbb{R}
\label{eq:IL}	
\end{equation}
\section{PROPOSED SOLUTION}
\subsection{PSO algorithm}
PSO is a population based optimization technique, where the population is called a swarm. Each particle represents a possible solution to the optimization. During each iteration each particle accelerates in the direction of its own personal best solution found so far, as well as in the direction of the global best position discovered so far by any of the particles in the swarm. This means that if a particle discovers a promising new solution, all the other particles will move closer to it, exploring the region more thoroughly in the process \cite{CPSO}.

In a $n$-dimensional search space, $S\subseteq \mathbb{R}^n$, assume that the swarm consists of $N$ particles. The $i$-th particle is in effect an $n$-dimensional vector $x_i=(x_{i1},x_{i2},\cdots,x_{in})\in S$. The velocity of this particle is also a $n$-dimensional vector $v=(v_{i1},v_{i2},\cdots,v_{in})\in S$. The best previous position visited by the $i$-th particle is a point in $S$, denoted as $p_i=(p_{i1},p_{i2},\cdots,p_{in})$. Let $g$ be the index of the particle that attained the best previous position among the entire swarm, and $t$ be the iteration counter. Then in PSO, the swarm is manipulated according to the following update equations \cite{SAVPSO}

\begin{equation}
\begin{aligned}
v_{id}(t+1) = & \omega |p_{i^{'}d}(t)-p_{id}(t)|\text{sign}(v_{id}(t)) \\
              & +r(p_{id}(t)-x_{id}(t))+(1-r)(p_{gd}(t)-x_{id}(t))\\      
\end{aligned}
\label{eq:PSAVPSO.8}	
\end{equation}
\begin{equation}
x_{id}(t+1) = x_{id}(t) + v_{id}(t+1)
\label{eq:PSAVPSO.9}	
\end{equation}
where $i = 1, 2, \cdots N$ is the particle's index, $d = 1, 2,\cdots, n$ indicates the particle's $d$-th component, $r \in U[0,1]$, $i^{'}\in intU[0,1]$, $\omega$ is a scaling parameter, and $\text{sign}(v_{id}(t))$ is the sign of $v_{id}(t)$.

The update equation of velocity (\ref{eq:PSAVPSO.8}) is slightly different from that usually used by PSO, this improved update rule highlights the exploration and exploitation abilities of the particles, which are adjusted with one parameter, $\omega$. If $\omega > 1$, the speed $\omega |p_{i^{'}d}(t)-p_{id}(t)|$ is expanded and thus the search scope of the swarm is enlarged, hence the exploration ability of the swarm is improved, but the convergence speed is lowered. If $\omega < 1$, then speed $\omega |p_{i^{'}d}(t)-p_{id}(t)|$ is reduced and thus the search scope of the swarm shrinks, the exploitation ability of the swarm is improved, and the algorithm converges fast but is prone to get trapped into local optimum. To obtain a good balance between exploitation and exploration, it should be reasonable to take $\omega =1$. Or alternatively, $\omega$ may be set to $\omega = cr_3$, where $c$ is a parameter and $r_3 \in U[0,1]$. Note that if $c = 2$, then $\omega = 2r3$, and the mean value of $\omega$ is 1; if $c < 2$, then the mean value of $\omega < 1$.
\subsection{CPSO algorithm}
In certain tasks, multiple agents need to coordinate their behavior to achieve a common goal, a powerful method is to coevolve them in separate subpopulations, and test together in the common task \cite{multiA}. 

Cooperative Coevolution (CC) consists in partitioning the search space by splitting the solution vectors into smaller vectors, then each of these smaller search spaces is searched by a separate mechanism (either serially or in parallel). In the case of large-scale continuous optimization, the subcomponents size depends on whether the problem is separable or not, which requires an analysis of the interdependence of variables that are involved in the optimization problem. Since the interdependence of variables in the IA problems is not yet analyzed, we opt in this study for the simplest form of the CPSO which adopts equally 1-D sized subcomponents for the whole optimization process.

The original PSO uses a population of $n$-dimensional vectors, in CPSO\_S (named here CPSO for simplicity), these vectors can be partitioned into $n$ swarms of 1-D vectors, each swarm representing a dimension of the original problem. Thereby, each swarm attempts to optimize a single component of the solution vector, essentially a 1-D optimization problem \cite{CPSO}. 

Notice that the function to be minimized, $f$, requires an $n$-dimensional vector as input, since each swarm represents only a single dimension of the search space, it is  possible to directly compute the fitness of the individuals of a single population considered in isolation.

\begin{figure}[H]  
\begin{flushleft}
\includegraphics[width=0.7\textwidth]{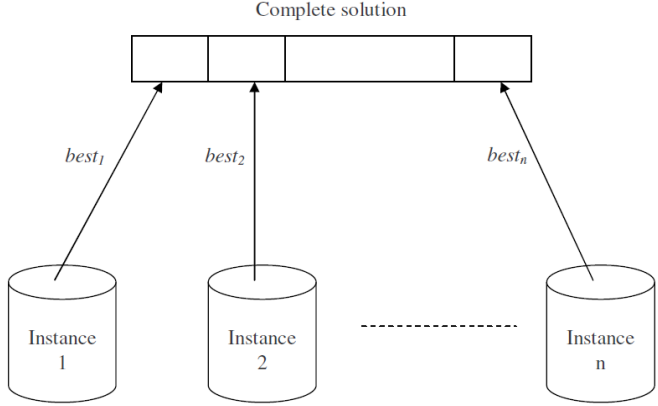}
  \caption{Cooperative approach based on explicit space decomposition.}
  \label{fig:image}
\end{flushleft}
\end{figure}

A \emph{context vector} is required to provide a suitable context in which the individuals of a population can be evaluated. The simplest scheme for constructing such a context vector is to take the global best particle from each of the $n$ swarms and concatenating them to form such an $n$-dimensional vector. To calculate the fitness for all particles in swarm $j$, the other $n-1$ components in the context vector are kept constant (with their values set to the global best particles from the other $n-1$ swarms), while the $j$th component of the context vector is replaced in turn by each particle from the $j$th swarm \cite{CPSO}. Fig. \ref{fig:image} illustrates illustrate the CC principle \cite{CABC}.

Table I presents the CPSO algorithm as introduced in \cite{CPSO}, $P_j.\mathbf{x_i}$ refers to the position of particle $i$ of
swarm $j$, which can therefore be substituted into the $j$th component of the context vector when needed. Each of the $n$ warms
now has a global best particle $P_{j}.\widehat{\mathbf{y}}$. The function $\mathbf{b}(j,z)$ returns an $n$-dimensional vector formed by concatenating all the global best vectors across all swarms, except for the $j$th component, which is replaced with $z$ where $z$ represents the position of any particle from swarm $P_j$.
\begin{table}[h!]
\begin{tabular}{l}
\hline
\textbf{Algorithm 1} The CPSO Algorithm\\
\hline
\textbf{1:~define}:\\
$\mathbf{b}(j,z) \equiv (P_1.\widehat{\mathbf{y}},P_2.\widehat{\mathbf{y}},\cdots,P_{j-1}.\widehat{\mathbf{y}},z,\cdots,P_{j+1}.\widehat{\mathbf{y}},\cdots, P_{n}.\widehat{\mathbf{y}})$\\
\textbf{2:}~Create and initialize $n$ one-dimensional PSOs: $P_j$, $j\in [1,\cdots,n]$\\
\textbf{3:}~\textbf{repeat:}\\
\textbf{4:}~~\textbf{for} each swarm $j\in [1,\cdots,n]$\\
\textbf{5:}~~~\textbf{for} each particle $i\in [1,\cdots,s]$\\
\textbf{6:}~~~~\textbf{if} $f(\mathbf{b}(j,P_j.\mathbf{x_i}))<f(\mathbf{b}(j,P_j.\mathbf{y_i})))$\\
\textbf{7:}~~~~~~\textbf{then} $P_j.\mathbf{y_i}=P_j.\mathbf{x_i}$\\
\textbf{8:}~~~~\textbf{if} $f(\mathbf{b}(j,P_j.\mathbf{y_i}))<f(\mathbf{b}(j,P_j.\widehat{\mathbf{y}})))$\\
\textbf{9:}~~~~~~\textbf{then} $P_j.\widehat{\mathbf{y}}=P_j.\mathbf{y_i}$\\
\textbf{10:}~~\textbf{endfor}\\
\textbf{11:}~~Perfom PSO updates on $P_j$ using equations (\ref{eq:PSAVPSO.8}-\ref{eq:PSAVPSO.9})\\
\textbf{12:}~\textbf{endfor}\\
\textbf{13:}~\textbf{until} stopping condition is true\\
\hline
\end{tabular}
\caption{Pseudocode for CPSO algorithm}
\end{table}

In this study we propose to use PSO and CPSO algorithms to minimize the IL given by equation (\ref{eq:IL}).
\section{Simulation results}
To verify the effectiveness of the proposed approach, we will perform a set of experiments conducted on six $K$-User MIMO IC scenarios. 

As stated in Section II, let's consider a $K$ user MIMO IC consisting $K$ transmitter receiver pairs, the transmitters and receivers are equipped with $M=N=5$ antennas each, and every transmitter aims to send $d=2$ data streams to its corresponding receiver. These "reasonable" settings are chosen according to our main reference \cite{P3} to allow a comparison between the two approaches. This scenario is tested with an increasing $K$ of 3, 5, 7, 9, 11 and 13, this complexity is somewhat excessive, but this will enable to emphasize the effectiveness of the proposed solution.

The number of complex elements of the decoding matrix $\mathbf{U}$ added to the number of complex elements of the precoding matrix $\mathbf{V}$ gives us the total number of the decision variables, in our case this sum is equal to $(K \times N \times d)+(K \times M \times d)$.

However, expression (\ref{eq:IL}) which gives the function to be minimized, $f(\mathbf{x})$, shows that $f(\mathbf{x})$ is a real-valued function of complex valued variables, i.e. not homomorphic \cite{wirtinger}, the optimization of this type of functions uses the so-called Wirtinger calculus \cite{wirtinger2}. Simply put, when optimizing real functions of one or more complex variables, we consider each complex variable as two real independent variables, the real part and the imaginary part. Thus optimization can be done as for multidimensional real functions \cite{wirtinger}. According to this, the total number of the decision variables involved  in our optimization process is equal to $2 \times (K \times N \times d)+2\times (K \times M \times d)$, Table II gives the dimension of each tested scenario.

At first, experiments are conducted using PSO algorithm, the entries of the MIMO channels are independent and identically distributed complex Gaussian variables with zero mean and unit variance, the swarm size is set to 100 and $\omega = 3$. Several runs was executed before fixing the swarm size and $\omega$, afterward each scenario was optimized over 10 independents runs.

\begin{figure}[H]
\begin{flushleft}
  \includegraphics[width=0.9\textwidth]{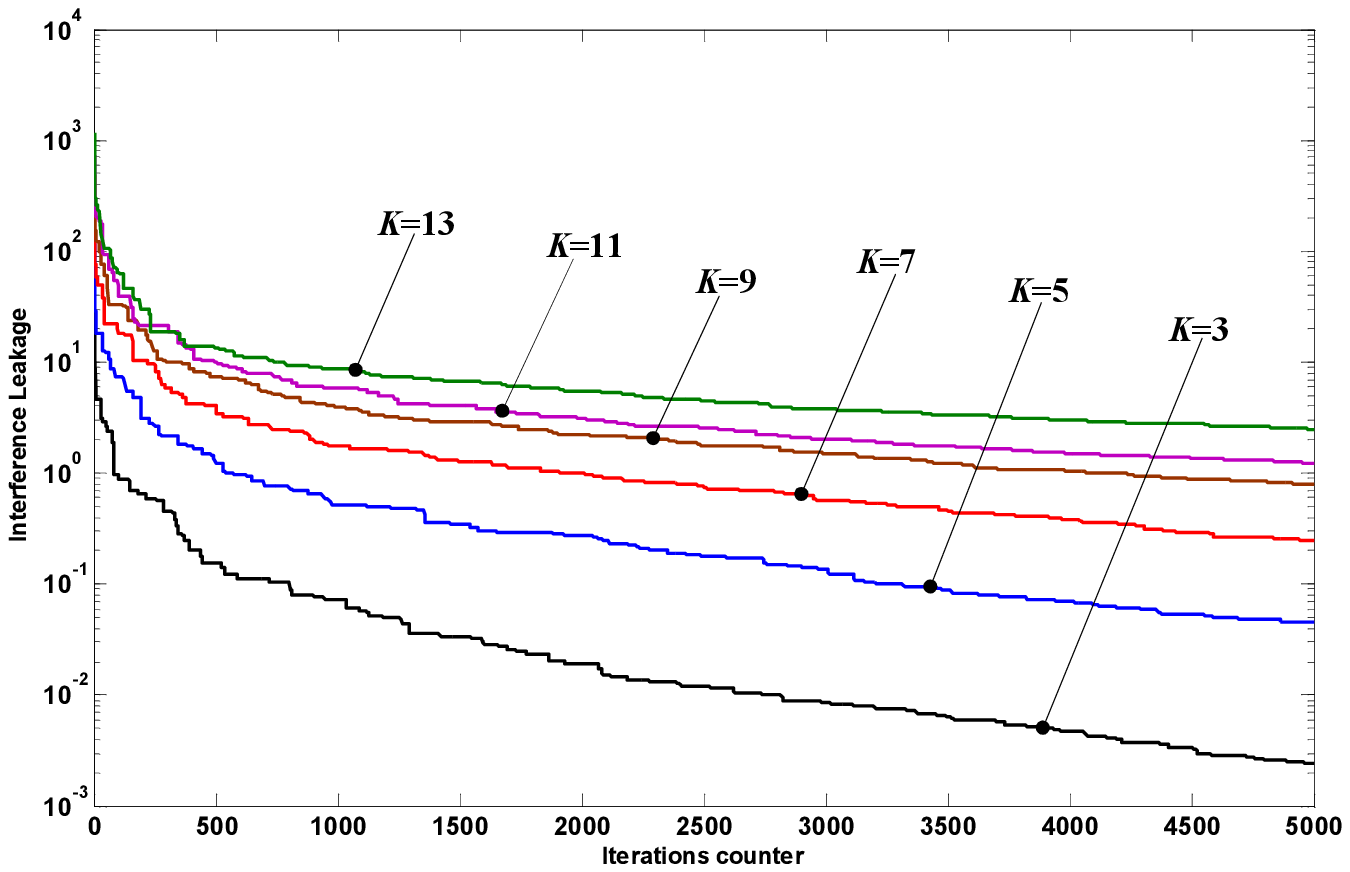}
  \caption{PSO simulation results.}
  \label{fig:pso}
\end{flushleft}
\end{figure}

Fig. \ref{fig:pso} depicts the evolution of the Interference Leakage with the iterations counter, the slow convergence is evident, and the calculation was stopped after 5000 iterations because beyond this value the optimization process becomes meaningless. Moreover, Table II shows more precisely that the level of the IL is not low enough for a signal-to-noise ratio (SNR) regime where IA is significant. 

\begin{figure}[H]
\begin{flushleft}
  \includegraphics[width=0.9\textwidth]{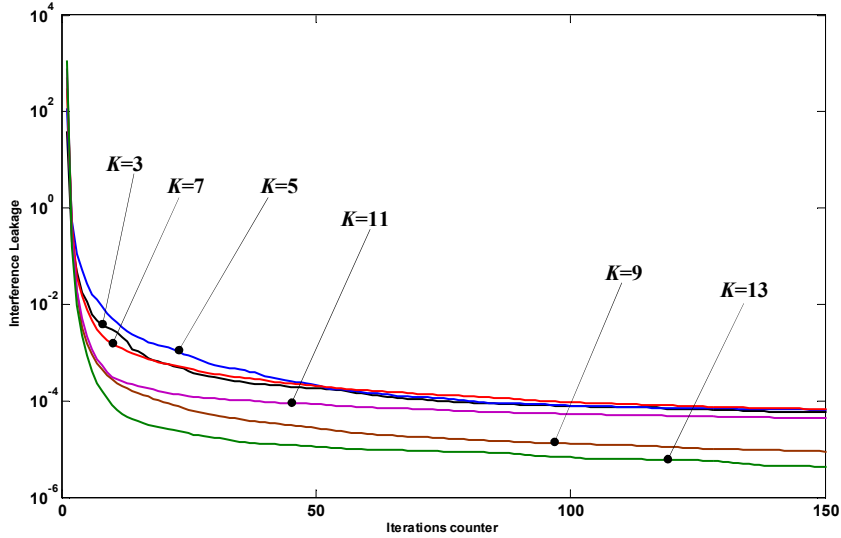}
  \caption{CPSO simulation results.}
  \label{fig:cpso}
\end{flushleft}
\end{figure}

In a second stage, the same scenarios were optimized with the CPSO algorithm using a number of swarms equal to the dimension of each problem (see table II), all the swarms have a size of 50 and $\omega$ was fixed to $10^-3$, these values were chosen after several trials. The convergence plot shown in Fig. \ref{fig:cpso} illustrates a much better convergence behavior. According to Table II, the IL has reached satisfactory values, these values are also close to the one adopted by reference \cite{P3} which is of about $10^{-5}$.
\begin{table}[H]
\begin{flushleft}
\begin{tabular}{|l|l|l|l|}
\hline
$K$ & Dimension & IL (PSO) & IL (CPSO) \\
\hline
3 & 120 &  0.0024 & $5.8298\times 10^{-5}$ \\
\hline
5 & 200 & 0.0448 &  $6.4305\times 10^{-5}$ \\
\hline
7 & 280 & 0.2479 & $6.6484\times 10^{-5}$ \\
\hline
9 & 360 & 0.7851 &  $8.7648\times 10^{-6}$ \\
\hline
11 & 440 & 1.2327 & $4.3054\times 10^{-5}$ \\
\hline
13 & 520 & 2.4852 & $4.2649\times 10^{-6}$ \\
\hline
\end{tabular}
\end{flushleft}
\caption{Simutation results}
\end{table}

\section{Conclusion}
In this paper, a CPSO based AI solution is proposed for the $K$-user MIMO IC. The CC approach was privileged because the optimization problem is of large scale, which has significantly improved the effectiveness of the metheuristic.
We can say that the metheuristic based IA solutions can be a serious alternative to the algebraic IA methods, because convergence is less constrained with hard algebraic assumptions. However, this first attempt should be improved especially by analyzing the interdependence of the decision variables this will reduce the computation time and improve the convergence rate.


\end{document}